\documentclass[twocolumn,letter]{jpsj2} 
\title{Superconducting Transition in the $\beta$-Pyrochlore AOs$_2$O$_6$ (A=Cs, Rb, K) \\ under Pressure}

\author{Kiyotaka \textsc{Miyoshi}, Yuta \textsc{Takaichi}, Yusuke \textsc{Takamatsu}, Motonobu \textsc{Miura} and Jun \textsc{Takeuchi}}

\inst{%
Department of Material Science, Shimane University, Matsue 690-8504
}

\abst{Pressure dependence of superconducting transition temperature $T_{\rm c}$ 
has been determined through the DC magnetic measurements under pressure up to $P$=10 GPa 
for $\beta$-pyrochlore oxides AOs$_2$O$_6$ with A=Cs ($T_{\rm c}$=3.3 K), Rb (6.3 K) and K (9.6 K). 
Both for A=Rb and Cs, $T_{\rm c}$ increases with increasing $P$ and shows 
a saturation at $T_{\rm cm}$$\sim$8.8 K, which is considered as the upper limit of $T_{\rm c}$ inherent in AOs$_2$O$_6$. 
In contrast, the $T_{\rm c}-P$ curve for KOs$_2$O$_6$ shows a sharp maximum of $\sim$10 K at $P$$\sim$0.5 GPa, 
and $T_{\rm c}$ is higher than $T_{\rm cm}$ for 0$\leq$$P$$\leq$1.5GPa, 
suggesting the enhanced superconductivity due to the rattling of K ions.

}

\kword{$\beta$-pyrochlore, high pressure, KOs$_2$O$_6$, RbOs$_2$O$_6$, CsOs$_2$O$_6$, superconductivity, 
DC magnetization, diamond anvil cell}

\begin{document}
\maketitle

In recent years, there has been a great deal of interest 
in pyrochlore or spinel oxides, in which a variety of remarkable 
low-temperature properties {\it e.g.}, metal-insulator\cite{shimakawa,cdos,miyoshi,yamamoto} or 
superconducting transition,\cite{hanawa,sakai,johnston} the exotic magnetic ground state 
such as spin ice\cite{harris,bramwell} and heavy mass Fermi-liquid behavior,\cite{kondo} 
have been found, exploring new types of electronic behavior on the geometrically frustrated lattice. 
Among these, the recent discovery of $\beta$-pyrochlore superconductor 
AOs$_2$O$_6$ (A=Cs,\cite{yonezawa1} Rb\cite{yonezawa2,bruhwiler1} and K\cite{yonezawa3}) 
may stimulate renewed interest in superconductivity 
on geometrically frustrated structures, as in LiTi$_2$O$_4$\cite{johnston} 
and Cd$_2$Re$_2$O$_7$.\cite{hanawa,sakai} 
Indeed, the superconductivity in AOs$_2$O$_6$ has been the subject of intensive research and 
several experiments suggest the conventional s-wave 
superconductivity.\cite{khasanov,magishi,bruhwiler2,kasahara,shimono,shimojima} 

One of the characteristic features of AOs$_2$O$_6$ is the anharmonic "rattling" motion 
of A ions in an oversized cage of Os-O network,\cite{kunes1} which has been experimentally 
inferred from the specific heat data 
showing the existence of low frequency Einstein mode contribution,\cite{hiroi1,bruhwiler3} 
and also demonstrated by the recent NMR experiments.\cite{yoshida}    
Especially for KOs$_2$O$_6$, the superconducting transition at $T_{\rm c}$=9.6 K is followed by 
a second transition concerning to the rattling freedom of K ion at $T_{\rm p}$$\sim$7.5 K, 
and a field-independent specific heat anomaly of first order transition 
has been observed at $T_{\rm p}$.\cite{hiroi2,hiroi3}
The measurements of electrical resistivity $\rho$ as a function of temperature $T$ in high magnetic fields have revealed 
that $\rho$($T$) changes the curvature from a concave-downward one at high temperatures to a  
Fermi-liquid behavior $\rho$($T$)$\propto$$AT^2$ below $T_{\rm p}$, 
suggesting that the electron-rattling phonon scattering disappears below $T_{\rm p}$.\cite{hiroi4,hiroi5} 
In RbOs$_2$O$_6$ and CsOs$_2$O$_6$, such a crossover of $\rho$($T$) has been observed at $T^{\rm *}$$\sim$15 K and $\sim$20 K, 
respectively, suggesting that the rattling motion is frozen below $T^{\rm *}$.\cite{hiroi5} 
Moreover, a strong coupling between the rattling motion of K ions and quasiparticles has been 
evidenced through the recent microwave penetration depth study, where it is suggested that the 
rattling phonons help to enhance superconductivity and K sites freeze to an ordered state below $T_{\rm p}$.\cite{shimono}  
Also, the ordering of K ions at $T_{\rm p}$ has been suggested from the theoretical side.\cite{kunes2}  
Thus, there exists an intriguing possibility that novel superconductivity enhanced by the rattling 
phonons is realized between $T_{\rm p}$ and $T_{\rm c}$ in KOs$_2$O$_6$. 

To gain more insight into the mechanism of the superconductivity in KOs$_2$O$_6$, 
it is important to clarify the intrinsic $T_{\rm c}$ inherent in AOs$_2$O$_6$ 
by the application of physical pressure $P$, 
which elevates $T_{\rm c}$ as expected from the relation between $T_{\rm c}$ and 
lattice constant ranging from 10.099 \AA \ (A=K) to 10.148 \AA \ (A=Cs), 
and examine whether or not $T_{\rm c}$ for KOs$_2$O$_6$ is really 
enhanced compared with that for other members. 
In this view, to establish the $T_{\rm c}$$-$$P$ relations for AOs$_2$O$_6$ is 
of significant importance.  
In the present work, we have performed DC magnetization measurements for AOs$_2$O$_6$ (A=Cs, Rb, K) 
under high pressure up to $P$=10 GPa and determined the precise $T_{\rm c}$$-$$P$ relations. 
It has been found that the $T_{\rm c}$$-$$P$ curve both for A=Rb and Cs saturates 
at $T_{\rm cm}$=8.8 K, which is thought of to be 
the upper limit of $T_{\rm c}$ inherent in AOs$_2$O$_6$. 
On the other hand, $T_{\rm c}$ for A=K is higher than $T_{\rm cm}$=8.8 K for 0$\leq$$P$$\leq$1.5GPa, 
suggesting the enhanced superconductivity in KOs$_2$O$_6$.  

The polycrystalline samples of AOs$_2$O$_6$ (A=Cs, Rb, K) used in this study were synthesized 
by the solid-state reaction technique similar to that described in the literatures.\cite{yonezawa1,yonezawa2,yonezawa3} 
For the products, powder X-ray diffraction measurements were examined and most of the diffraction peaks 
were indexed as a pyrochlore structure with a lattice parameter consistent with earlier results.\cite{yonezawa1,yonezawa2,yonezawa3} 
A small amount of OsO$_2$ was also detected as an impurity. 
$T_{\rm c}$ at ambient pressure for all samples was confirmed to 
agree with literatures.\cite{yonezawa1,yonezawa2,yonezawa3}
For the magnetic measurements under high pressure up to 10 GPa, a miniature diamond anvil cell (DAC) with an outer 
diameter of 8 mm was used to generate high pressure and 
combined with a sample rod of a commercial SQUID magnetometer. 
The details of the DAC are given in elsewhere.\cite{mito}
The AOs$_2$O$_6$ sample was loaded into the gasket hole together 
with a small piece of high purity lead (Pb) to realize $in$ $situ$ observation of pressure
by determining the pressure from the $T_{\rm c}$ shift of Pb. 
Daphne oil 7373 was used as a pressure transmitting medium. 

To investigate the $T_{\rm c}$$-$$P$ relations,  
zero-field-cooled DC magnetization $M$ versus temperature data were collected at each pressure and 
the diamagnetic onset temperature was defined as $T_{\rm c}$. 
For all samples, the $M$$-$$T$ curve at ambient pressure was confirmed to remain unchanged after pressure cycling.  
First, we show the $M$$-$$T$ curves for KOs$_2$O$_6$ under various pressures in Fig. 1. 
In the figure, the $M$$-$$T$ curve at ambient pressure exhibits a sudden decrease 
indicating the superconducting transition in KOs$_2$O$_6$ at $T_{\rm c}$$\sim$9.6 K in addition to a sharp drop at $\sim$7 K 
corresponding to the diamagnetic onset of Pb for the pressure calibration. 
Both of the $M$$-$$T$ curves at $P$=0.35 and 0.53 GPa shows a diamagnetic onset at $T_{\rm c}$$\sim$10 K, 
higher than $T_{\rm c}$ at ambient pressure. However, the $M$$-$$T$ curve at $P$=1.1 GPa 
indicates a lower $T_{\rm c}$ of $\sim$9.4 K, and $T_{\rm c}$ appears to be 
systematically decreased as the pressure is increased as $P$$\geq$0.53 GPa in Fig. 1. These results indicate that 
$T_{\rm c}$ for KOs$_2$O$_6$ exhibits a maximum of $\sim$10 K at $P$$\simeq$0.4$-$0.5 GPa, 
consistent with the results of the magnetic measurements under pressure $P$$\leq$1.2 GPa 
by Muramatsu $et$ $al$.\cite{muramatsu1} 
The $M$$-$$T$ curves at $H$=20 Oe does not show an anomaly corresponding to the transition at $T_{\rm P}$, different from 
those at $H$$\geq$1 T.\cite{hiroi3,hiroi4}  

\begin{figure}[t]
\begin{center}
\includegraphics[width=7.7cm]{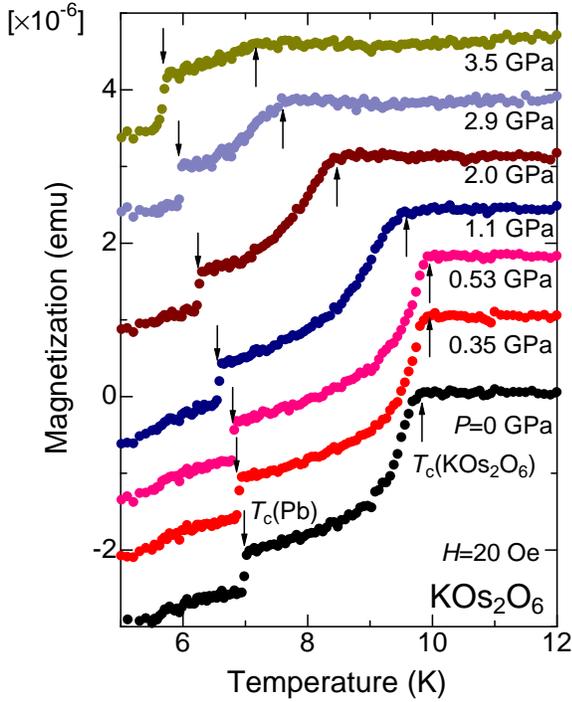}
\end{center}
\caption{(Color on line) Temperature dependence of zero-field-cooled DC magnetization 
for KOs$_2$O$_6$ under various pressures. 
The positions of $T_{\rm c}$ for KOs$_2$O$_6$ and Pb are indicated by upward and downward arrows, respectively. 
The data are intentionally shifted along the vertical axis for clarity.}
\label{f1}
\end{figure}
Next, we show typical results of the $M$$-$$T$ measurements under pressure 
for RbOs$_2$O$_6$ and CsOs$_2$O$_6$ in Figs. 2 and 3, respectively. 
As seen in Fig. 2, the $M$$-$$T$ curve at $P$=0 GPa for RbOs$_2$O$_6$ exhibits a two-step decrease 
with decreasing temperature corresponding to the superconducting transition in Pb at $\sim$7 K 
and RbOs$_2$O$_6$ at $\sim$6.3 K. $T_{\rm c}$ for RbOs$_2$O$_6$ shifts toward higher temperature side
as increasing $P$ and reaches $\sim$8.8 K at $P$=3.0 GPa. For the further increase of pressure, 
$T_{\rm c}$ becomes pressure independent but appears to be decreased for $P$$>$4.7 GPa. 
For $P$$\geq$4.7 GPa, the decreasing rate of $M$ below $T_{\rm c}$ becomes smaller, suggesting that 
the superconducting transition is more gradual in connection with the decrease of $T_{\rm c}$. 
On the other hand, the superconducting transition for CsOs$_2$O$_6$ is almost 
overlapped with that for Pb at $P$=2.7 GPa in Fig. 3. 
In the figure, $T_{\rm c}$ appears to be monotonically increased by the application of pressure 
and is achieved to $\sim$8.8 K at $P$=6.2 GPa 
but appears to be independent of pressure for 6.2 GPa$<$$P$$<$8.2 GPa. 
Since the diamagnetic onset was unclear due to the broadening of the superconducting transition for $P$$>$8.2 GPa, 
it is uncertain whether or not $T_{\rm c}$ for CsOs$_2$O$_6$ is lowered from $\sim$8.8 K 
under high pressure as in RbOs$_2$O$_6$. 

\begin{figure}[t]
\begin{center}
\includegraphics[width=7.7cm]{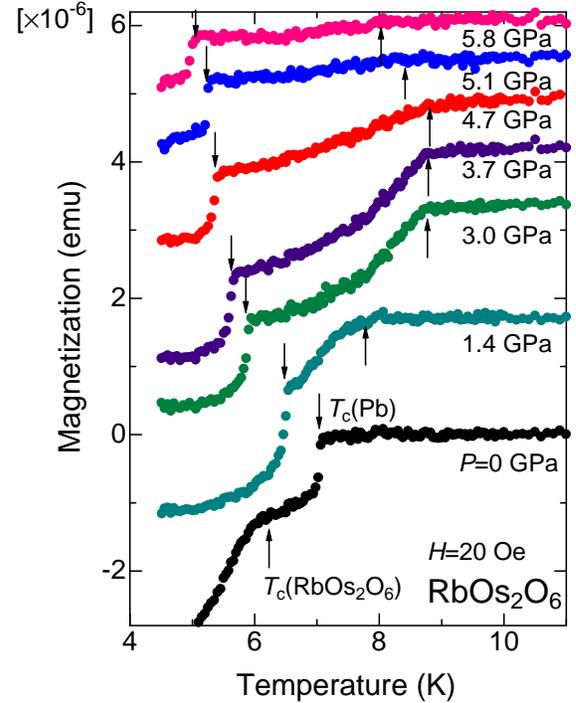}
\end{center}
\caption{(Color on line) Temperature dependence of zero-field-cooled DC magnetization 
for RbOs$_2$O$_6$ under various pressures. 
The positions of $T_{\rm c}$ for RbOs$_2$O$_6$ and Pb are indicated by upward and downward arrows, respectively. 
The data are intentionally shifted along the vertical axis for clarity.}
\label{f1}
\end{figure}
\begin{figure}[t]
\begin{center}
\includegraphics[width=7.7cm]{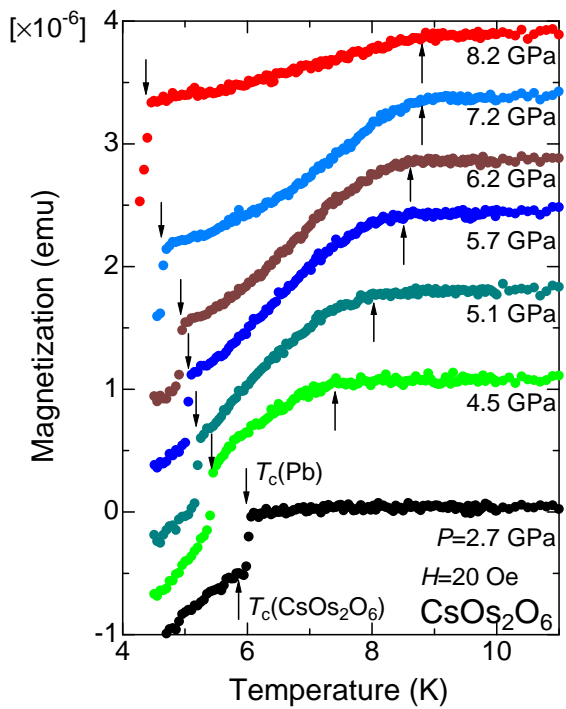}
\end{center}
\caption{(Color on line) Temperature dependence of zero-field-cooled DC magnetization 
for CsOs$_2$O$_6$ under various pressures. 
The positions of $T_{\rm c}$ for CsOs$_2$O$_6$ and Pb are indicated by upward and downward arrows, respectively. 
The data are intentionally shifted along the vertical axis for clarity.}
\label{f1}
\end{figure}
In Fig. 4(a), $T_{\rm c}$ versus pressure data for AOs$_2$O$_6$ are summarized. 
In the figure, the $T_{\rm c}$$-$$P$ curve for RbOs$_2$O$_6$ displays a 
monotonic increase for $P$$<$3 GPa and a plateau between $P$$\sim$3 GPa and 5 GPa. 
The $T_{\rm c}$$-$$P$ curve for CsOs$_2$O$_6$ also 
shows a monotonic increase and a saturation for $P$$>$6 GPa. 
The $T_{\rm c}$$-$$P$ curve both for A=Rb and Cs saturates at the same temperature $T_{\rm cm}$=8.8 K. 
The increasing rate of the $T_{\rm c}$$-$$P$ curve for $T_{\rm c}$$<$$T_{\rm cm}$ is 
0.9$-$1.0 K$/$GPa for both materials. The shapes of the $T_{\rm c}$$-$$P$ curves are 
similar to each other although a decrease 
from $T_{\rm cm}$ at high pressure was not observed for CsOs$_2$O$_6$. 
In striking contrast, the $T_{\rm c}$$-$$P$ curve for KOs$_2$O$_6$ exhibits a 
sharp maximum of $\sim$10 K which is nearly 1 K higher than $T_{\rm cm}$. 
The $T_{\rm c}$$-$$P$ curves previously reported by Muramatsu $et$ $al$.\cite{muramatsu2} 
are also described in the figure for comparison. 
They have determined $T_{\rm c}$ through the $\rho$$-$$T$ measurements 
for polycrystalline samples under pressure up to 10 GPa using a cubic anvil press except for 
$P$$\leq$1.2 GPa. They have adopted the diamagnetic onset of 
the $M$$-$$T$ curve as $T_{\rm c}$ for $P$$\leq$1.2 GPa.\cite{muramatsu1}  
For KOs$_2$O$_6$, the $T_{\rm c}$$-$$P$ curve coincides with that by Muramatsu $et$ $al$. for 
$P$$<$2 GPa but not for $P$$>$2 GPa. 
The $T_{\rm c}$$-$$P$ curve based on the $\rho$$-$$T$ measurements for RbOs$_2$O$_6$ (CsOs$_2$O$_6$) 
displays a broad maximum of $\sim$8 K at $P$=2 GPa ($\sim$7.5 K at $P$=6 GPa), 
different from that based on the $M$$-$$T$ measurements, but is coincident 
with each other for $P$$<$2 GPa ($P$$<$4 GPa). 
The inconsistency is mainly coming from the difficulty to determine $T_{\rm c}$ from 
the $\rho$$-$$T$ curve, which shows a resistive drop over wide temperature range of $\Delta$$T$=1$-$5 K 
under high pressure.\cite{muramatsu2} 
Muramatsu $et$ $al$. has determined $T_{\rm c}$ as the midpoint of the resistive drop. 
The onset of the resistive drop would be rather close to the diamagnetic onset in the present work.   
\begin{figure}[t]
\begin{center}
\includegraphics[width=8cm]{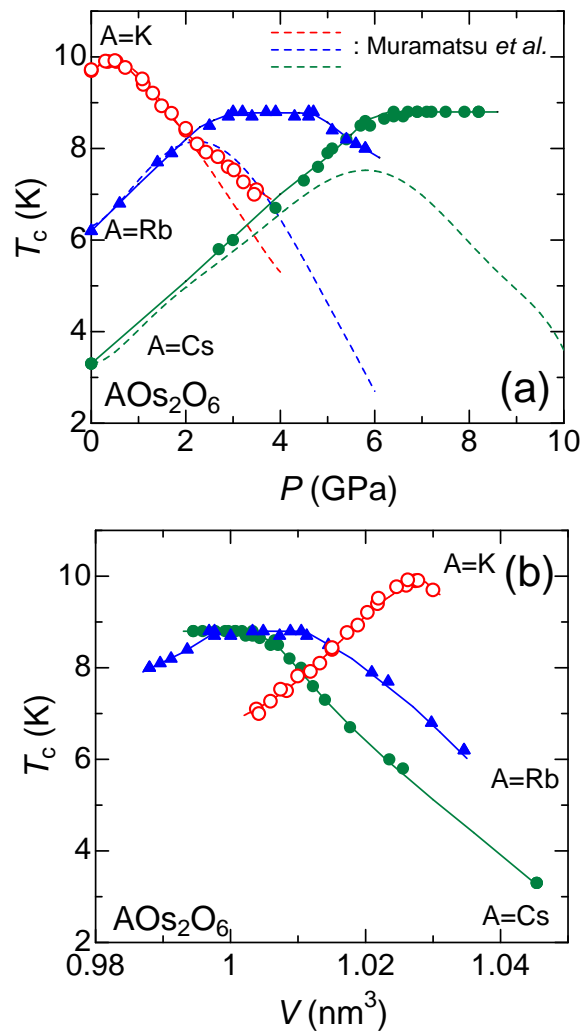}
\end{center}
\caption{(Color on line) (a) Pressure ($P$) dependence of superconducting transition temperature $T_{\rm c}$ for 
AOs$_2$O$_6$ (A=Cs, Rb, K). The solid lines are guides for the eyes. 
The dashed lines represent the $T_{\rm c}$$-$$P$ curves determined from the 
electrical resistivity measurements by Muramatsu $et$ $al$. reproduced from ref. 32. 
(b) Plots of $T_{\rm c}$ versus unit cell volume ($V$) for AOs$_2$O$_6$ (A=Cs, Rb, K) obtained 
by transforming the $T_{\rm c}$$-$$P$ data using the relation $\Delta$$V$$/$$V_{0}$=$-$$\alpha$$P$+$\beta$$P^2$ 
in ref. 28. The solid lines are guides for the eyes. 
}
\label{f1}
\end{figure}

As seen in Fig. 4(a), our magnetic measurements under pressure reveal the characteristic $T_{\rm c}$$-$$P$ curves, 
although the evolution of the overall band structures 
for the physical compression is known to be rather small.\cite{saniz1} 
Particularly for KOs$_2$O$_6$, the $T_{\rm c}$$-$$P$ curve 
takes a maximum of $\sim$10 K higher than those for other members ($T_{\rm cm}$=8.8 K)
indicating the enhanced superconductivity or the extension of the superconducting region 
in the $T$$-$$P$ phase diagram. 
An enhancement of superconductivity in KOs$_2$O$_6$ is also suggested  
by Shimono $et$ $al$. through the measurements of microwave surface impedance.\cite{shimono} 
They have shown that the temperature dependence of superfluid density 
exhibits a step-like change near $T_{\rm p}$ and extrapolates to zero at $T_{\rm 0}$=8.7 K. 
$T_{\rm 0}$ is the effective $T_{\rm c}$ estimated from the behavior below $T_{\rm p}$ where the rattling motion freezes, 
and $T_{\rm c}$(=9.6 K) is surely higher than $T_{\rm 0}$, suggesting the enhanced superconductivity and the importance of the 
rattling motion for the enhancement. 
In a similar context, the recent Hall-sensor magnetometory has revealed that the lower critical field versus temperature data 
below $T_{\rm p}$ extraporates to zero at 9.2 K, lower than $T_{\rm c}$, suggesting that 
the superconducting gap is reduced below $T_{\rm P}$.\cite{shibauchi}
Indeed, $T_{\rm 0}$(=8.7 K) is considered as the original $T_{\rm c}$ in the absence of the rattling of K ions, 
and is in good agreement with the upper limit of $T_{\rm c}$ for A=Rb and Cs ($T_{\rm cm}$=8.8 K) inferred from 
the saturation of the $T_{\rm c}$$-$$P$ curve. 
This suggests that the upper limit of $T_{\rm c}$ 
is basically $T_{\rm cm}$=8.8 K in AOs$_2$O$_6$ superconductor but $T_{\rm c}$ for KOs$_2$O$_6$ 
is markedly enhanced from the upper limit probably due to the rattling motion of K ions. 
It is expected that, in case of the absence of the rattling motion in KOs$_2$O$_6$, 
the maximum of $T_{\rm c}$ would be limited to 8.8 K, and then 
a plateau of the $T_{\rm c}$$-$$P$ curve would 
be realized for 0$\leq$$P$$\leq$1.5 GPa.   

In Fig. 4(b), $T_{\rm c}$ versus unit cell volume $V$ data are plotted. We obtained 
the $T_{\rm c}$$-$$V$ data by transforming the $T_{\rm c}$$-$$P$ data 
using the relation $\Delta$$V$$/$$V_{0}$=$-$$\alpha$$P$+$\beta$$P^2$, 
where $\alpha$=0.00726 (GPa)$^{-1}$ and $\beta$=0 for KOs$_2$O$_6$, 
$\alpha$=0.00777 (GPa)$^{-1}$ and $\beta$=0 for RbOs$_2$O$_6$, 
and $\alpha$=0.00756 (GPa)$^{-1}$ and $\beta$=0.000198 (GPa)$^{-2}$ for CsOs$_2$O$_6$.\cite{hiroi5} 
Since $T_{\rm c}$ systematically changes with A ion radius or lattice constant in AOs$_2$O$_6$, 
one may expect that the $T_{\rm c}$ versus cell volume data for all members 
are interpolated by a universal curve indicating that $T_{\rm c}$ is determined by lattice constant 
independently on whether the lattice is compressed by chemical or physical pressure, 
such as the ferromagnetic transition temperature in R$_2$Mo$_2$O$_7$ pyrochlore (R=Nd-Gd).\cite{miyoshi2} 
As seen in Fig. 4(b), the $T_{\rm c}$$-$$V$ curves are however different from each other even between A=Rb and Cs. 
The increasing rate of $T_{\rm c}$ for the chemical compression $|$$dT_{\rm c}/dV$$|$ is 
estimated to be $\sim$2.7$\times$10$^2$ K$/$nm$^3$ considering 
the value of $T_{\rm c}$ and $V$ for A=Cs and Rb at ambient pressure, and 
is much larger than that for the physical one for A=Cs and Rb ($\sim$1.2$\times$10$^2$ K$/$nm$^3$) in Fig. 4(b). 
This is probably due to the difference between the modifications 
of local lattice structure induced by chemical and physical pressure. 
The relation between $T_{\rm c}$ and the detailed crystal structure 
under high pressure should be clarified for each system to 
advance towards further understanding of the mechanism of the superconductivity. 

In summary, it has been found that the $T_{\rm c}$$-$$P$ curve saturates with increasing $P$ reaching 
$T_{\rm cm}$=8.8 K both for A=Rb and Cs, while that for KOs$_2$O$_6$ exhibits a sharp maximum of $\sim$10 K 
at $P$$\sim$0.5 GPa. $T_{\rm cm}$ is thought of to be the upper limit of $T_{\rm c}$ inherent in AOs$_2$O$_6$  
superconductor. For KOs$_2$O$_6$, the superconducting region 
is extended in the $T$$-$$P$ phase diagram for 0$\leq$$P$$\leq$1.5GPa.
Since the rattling motion does not survive at low temperature below $T_{\rm cm}$ for A=Rb and Cs, 
it is inferred that the rattling of K ions plays a crucial role for 
the enhanced superconductivity in KOs$_2$O$_6$. 

\acknowledgements
This work is financially supported by a Grant-in-Aid for 
Scientific Research (No. 17740229) from the Japanese Ministry of Education, 
Culture, Sports, Science and Technology.

\end{document}